\documentclass[onecolumn,showpacs,amsfonts,amsmath,amssymb,floatfix,superscriptaddress]{revtex4}
\usepackage{url}
\usepackage{bm}		% bold math
\usepackage{graphicx}	% figures
\begin{document}

%\font\tenssf = cmss10
%\input{wasyfont}		% bold math

\title[The interior structure of rotating black holes 1]{The interior structure of rotating black holes 1. Concise derivation}

\author{Andrew J S Hamilton}
\email{Andrew.Hamilton@colorado.edu}	% revtex
%\ead{Andrew.Hamilton@colorado.edu}	% iopart
%\homepage{http://casa.colorado.edu/~ajsh/}
\affiliation{JILA, Box 440, U. Colorado, Boulder, CO 80309, USA}
\affiliation{Dept.\ Astrophysical \& Planetary Sciences,
U. Colorado, Boulder, CO 80309, USA}
\author{Gavin Polhemus}
\email{Gavin.Polhemus@colorado.edu}
\affiliation{JILA, Box 440, U. Colorado, Boulder, CO 80309, USA}

\newcommand{\simpropto}{\raisebox{-0.7ex}[1.5ex][0ex]{
		\begin{array}[b]{@{}c@{\;}} \propto \\
		[-1.8ex] \sim \end{array}}}

\newcommand{\dd}{d}
\newcommand{\ddsq}{\dd^2\mkern-1.5mu}
\newcommand{\ddd}{\dd^3\mkern-1.5mu}
\newcommand{\dddd}{\dd^4\mkern-1.5mu}
\newcommand{\DD}{D}
\newcommand{\ee}{e}
\newcommand{\im}{i}
\newcommand{\Ei}{{\rm Ei}}
\newcommand{\perpperp}{\perp\!\!\perp}
\newcommand{\ppartial}{\partial^2\mkern-1mu}
\newcommand{\nn}{\nonumber\\}

\newcommand{\diag}{{\rm diag}}
\newcommand{\expinf}{\xi}
\newcommand{\jel}{\text{\sl j}}
\newcommand{\Lz}{L}
\newcommand{\Msun}{{\rm M}_\odot}
\newcommand{\uel}{u}
\newcommand{\vel}{v}
\newcommand{\inn}{{\rm in}}
\newcommand{\out}{{\rm ou}}
\newcommand{\sep}{{\rm sep}}

\newcommand{\bg}{\bm{g}}
\newcommand{\bp}{\bm{p}}
\newcommand{\bv}{\bm{v}}
\newcommand{\bx}{\bm{x}}
\newcommand{\bgamma}{\bm{\gamma}}

\newcommand{\Apot}{{\cal A}}
\newcommand{\hatA}{\hat{A}}
\newcommand{\Br}{B}
\newcommand{\betar}{\lambda_r}
\newcommand{\Cx}{C_x}
\newcommand{\Cy}{C_y}
\newcommand{\Dx}{D_x}
\newcommand{\Dy}{D_y}
\newcommand{\Deltax}{\Delta_x}
\newcommand{\Deltaxinf}{\Delta_{x , {\rm inf}}}
\newcommand{\Deltaxev}{\Delta_{x , {\rm ev}}}
\newcommand{\Deltay}{\Delta_y}
\newcommand{\Er}{E}
\newcommand{\Fz}{{\tilde F}}
\newcommand{\starF}{\,{}^\ast\!F}
\newcommand{\Jx}{J_x}
\newcommand{\Jy}{J_y}
\newcommand{\Mass}{{\cal M}}
\newcommand{\Mbh}{M_\bullet}
\newcommand{\Mdot}{\dot{M}}
\newcommand{\KCarter}{{\cal K}}
\newcommand{\NUT}{{\cal N}}
\newcommand{\px}{p^x}
\newcommand{\Px}{P_x}
\newcommand{\Py}{P_y}
\newcommand{\QCarter}{{\cal Q}}
\newcommand{\Qelec}{Q}
\newcommand{\Qelecbh}{\Qelec_\bullet}
\newcommand{\Qmag}{{\cal Q}}
\newcommand{\DeltaQelec}{\Delta \Qelec}
\newcommand{\DeltaQmag}{\Delta \Qmag}
\newcommand{\rhosep}{\rho_{\rm s}}
\newcommand{\rhox}{\rho_x}
\newcommand{\rhoy}{\rho_y}
\newcommand{\Uinf}{U}
\newcommand{\Ur}{{\cal R}}
\newcommand{\Utheta}{\Theta}
\newcommand{\rc}{{\scriptstyle R}}
\newcommand{\tc}{{\scriptstyle T}}
\newcommand{\smallrc}{{\scriptscriptstyle R}}
\newcommand{\smalltc}{{\scriptscriptstyle T}}
\newcommand{\smallzero}{{\scriptscriptstyle 0}}
\newcommand{\Ux}{U_x}
\newcommand{\Uy}{U_y}
\newcommand{\xin}{x_{\rm in}}
\newcommand{\Xx}{X_x}
\newcommand{\Xy}{X_y}
\newcommand{\Yx}{Y_x}
\newcommand{\Yy}{Y_y}
\newcommand{\Zx}{Z_x}
\newcommand{\Zy}{Z_y}
\newcommand{\omegax}{\omega_x}
\newcommand{\omegay}{\omega_y}

\hyphenpenalty=3000

%--------------------
% FIG
\newcommand{\penrosekerrinflationfig}{
    \begin{figure}[t]
    \begin{center}
    \leavevmode
    \includegraphics[bb=186 264 463 529,scale=.9]{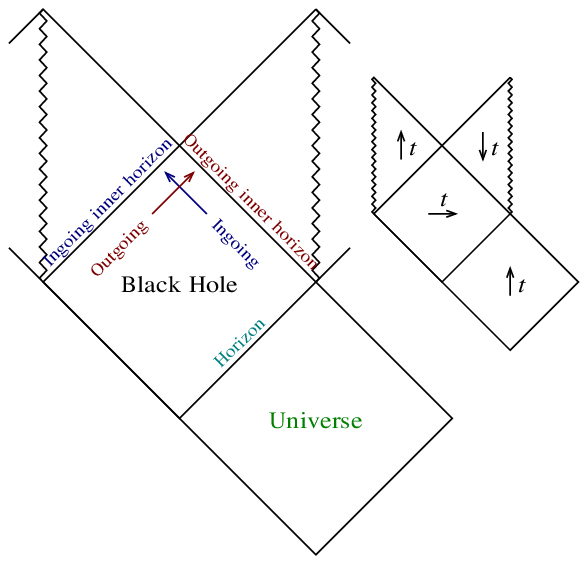}
    \caption[1]{
    \label{penrosekerrinflation}
Partial Penrose diagram illustrating why the Kerr geometry
is subject to the inflationary instability.
Ingoing and outgoing streams just outside the inner horizon
must pass through separate ingoing and outgoing inner horizons
into causally separated pieces of spacetime where the
timelike Kerr time coordinate $t$ goes in opposite directions.
To accomplish this, the ingoing and outgoing streams must
exceed the speed of light through each other,
which physically they cannot do.
In reality,
hyper-relativistic counter-streaming between the ingoing and outgoing streams
ignites and then drives the exponentially growing inflationary instability.
%The inflationary instability is driven by the pressure of the
%relativistic counter-streaming between ingoing and outgoing streams.
The inset shows the direction of coordinate time $t$
in the various regions.
Proper time of course always increases upward
in a Penrose diagram.
    }
    \end{center}
    \end{figure}
}

%--------------------
% FIG
\newcommand{\kerrrcontourfig}{
    \begin{figure}[tb]
    \begin{center}
    \leavevmode
    \includegraphics[bb=160 287 451 506,scale=.7]{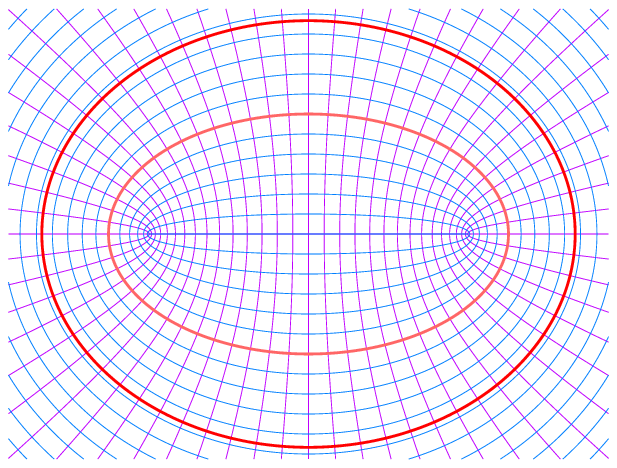}
    \includegraphics[bb=160 287 451 506,scale=.7]{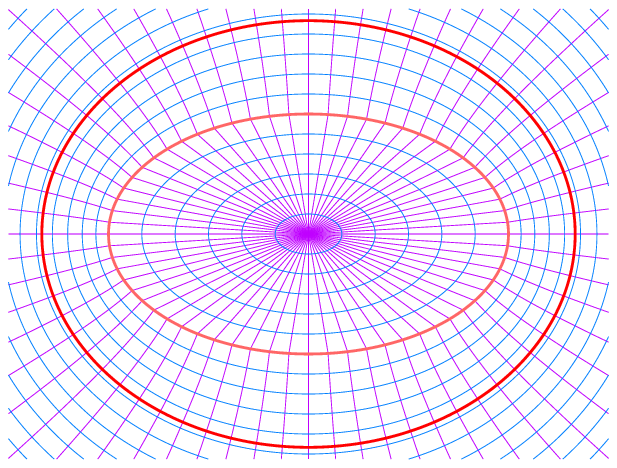}
    \caption[1]{
    \label{kerrrcontour}
Contours of constant radius $x$ and latitude $y$
in an uncharged black hole
with spin parameter
$a = 0.96 \Mbh$.
The thicker contours mark the outer and inner horizons.
The left panel depicts a Kerr black hole.
The right panel depicts a black hole of the kind
considered in the present series of papers,
which undergoes inflation just above the inner horizon, then collapses.
In the Kerr geometry,
surfaces of constant radius are confocal ellipsoids
in Boyer-Lindquist coordinates,
while surfaces of constant latitude are confocal hyperboloids,
with a ring singularity at their focus.
In the inflationary geometry,
the streaming energy density and pressure, and Weyl curvature,
inflate to exponentially huge values at (just above) the inner horizon,
which is destroyed.
In the conformally separable solutions presented here,
the geometry then collapses radially
to exponentially tiny size
without changing shape.
    }
    \end{center}
    \end{figure}
}

%--------------------
% FIG
\newcommand{\inflationaryhorizonfig}{
    \begin{figure}[tb]
    \begin{center}
    \leavevmode
    \includegraphics[bb=152 258 416 512,scale=.8]{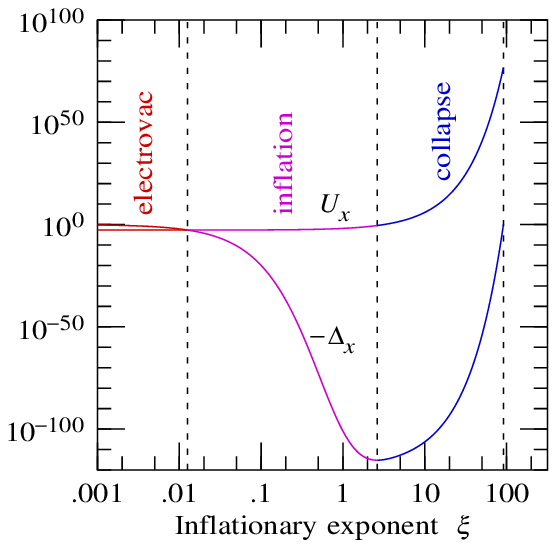}
    \includegraphics[bb=159 258 416 512,scale=.8]{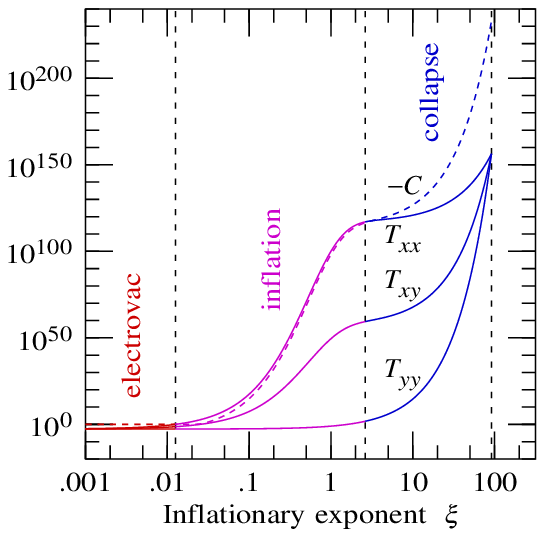}
    \caption[1]{
    \label{inflationaryhorizon}
Evolution of the geometry and energy-momenta
from electrovac through inflation and collapse.
(Left)
The parameter $\Ux \equiv - \Deltax \dd \xi / \dd x$
and
the horizon function $\Deltax$,
equations~(\protect\ref{UDxinf}),
as a function of the inflationary exponent $\expinf$,
for parameters
$\vel = 0.001$,
$\uel = 0.002$,
and $\Deltax^\prime = 1$
(the solutions in this paper apply
in the limit of tiny $\vel$ and $\uel$;
small finite values are adopted in this plot to avoid numerical overflow).
%In the actual solutions,
%the parameters $\uel$ and $\vel$ are to be considered as tiny;
%the values plotted here are ``large'' so that the evolutionary
%behaviour is discernible.
Inflation ignites as the horizon function $| \Deltax |$ decreases below $\Ux$.
Inflation ends as the absolute value of the horizon function
goes through a minimum,
and the geometry proceeds to collapse.
Once $| \Deltax | \gtrsim 1$,
the angular components of the collisionless streams
exceed their radial components, and the solution breaks down,
but this happens only after the geometry has collapsed to
exponentially tiny scale.
(Right)
The tetrad-frame
radial, radial-angular, and angular collisionless energy-momenta
$T_{xx} \propto \rho^{-2} \Ux / | \Deltax |$,
$T_{xy} \propto \rho^{-2} \Ux / \sqrt{| \Deltax |}$,
and
$T_{yy} \propto \rho^{-2} \Ux$.
The energy-momenta grow exponentially huge
despite their small initial values.
Indeed, the smaller the initial energy-momenta,
the faster and larger they grow.
The dashed line is minus the polar (real) spin-$0$
component of the Weyl curvature,
$- C \propto \rho^{-2} \Ux^2 / | \Deltax |$.
The axial (imaginary) spin-$0$ Weyl component
is comparable to $T_{yy}$.
    }
    \end{center}
    \end{figure}
}

%--------------------
% FIG
\newcommand{\rotinforbitfig}{
    \begin{figure}[tb]
    \begin{center}
    \leavevmode
    \includegraphics[bb=160 287 451 506,scale=1]{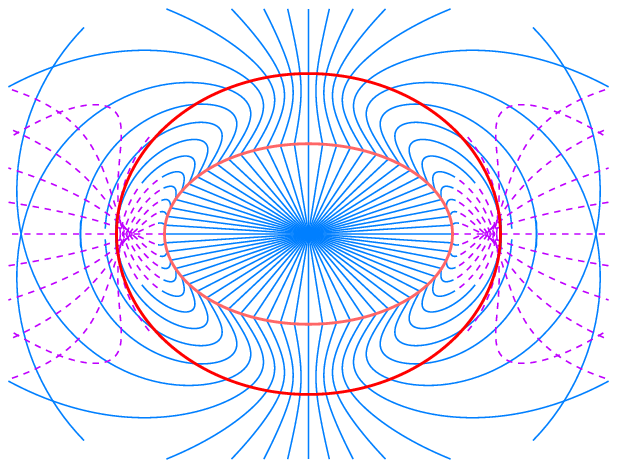}
    \caption[1]{
    \label{rotinforbit}
Angular flow pattern of freely-falling particles
that produces the conditions~(\protect\ref{ppm}) at the inner horizon required
if conformal separability is imposed to sub-dominant radial-angular order,
for an uncharged black hole
with spin parameter
$a = 0.96 \Mbh$.
The thicker contours mark the outer and inner horizons,
the latter being destroyed by inflation.
Ingoing and outgoing particles
fall along the same trajectories in the $x$-$y$ plane,
but have opposite motions in the azimuthal $\phi$ coordinate.
Trajectories near the equatorial plane
change from ingoing to outgoing, or vice versa,
inside the outer horizon;
the transition is marked by the lines
changing from dashed to solid.
The angular flow pattern cannot be achieved with collisionless streams
that fall from outside the outer horizon.
In the equatorial region, outgoing but not ingoing particles
can fall from outside the outer horizon,
while
in the polar region, ingoing but not outgoing particles
can fall from outside the outer horizon.
    }
    \end{center}
    \end{figure}
}

\begin{abstract}
This paper presents a concise derivation
of a new set of solutions
for the interior structure of accreting, rotating black holes.
The solutions are conformally stationary, axisymmetric,
and conformally separable.
Hyper-relativistic counter-streaming
between freely-falling collisionless ingoing and outgoing streams
leads to mass inflation at the inner horizon,
followed by collapse.
The solutions fail at an exponentially tiny radius,
where the rotational motion of the streams becomes comparable
to their radial motion.
%neutral or charged black holes derived in two companion papers.
The papers
%solve a longstanding problem,
provide
%for the first time
a fully nonlinear,
dynamical solution for the interior structure of a rotating black hole
from just above the inner horizon inward, down to a tiny scale.
\end{abstract}

	% Classical general relativity

\date{\today}

\maketitle

\section{Introduction}

Two companion technical papers
\cite{Hamilton:2010b,Hamilton:2010c},
hereafter Papers~2 and 3,
present
conformally stationary, axisymmetric, conformally separable
solutions for
the interior structure of a rotating black hole
that accretes a collisionless fluid,
undergoes inflation at its inner horizon,
and then collapses.
Paper~2 deals with uncharged black holes,
while Paper~3 generalizes to charged black holes.
%The claim is that these papers provide a solution to a problem that has
%remained outstanding since
%Kerr's 1963 \cite{Kerr:1963}
%discovery of the exterior geometry of a rotating black hole.
The purpose of the present paper is to
give an abbreviated derivation of the solution
for an uncharged black hole,
and to summarize the principal features of the solution.
A Mathematica notebook containing many details of the calculations
is at \cite{Hamilton:notebook}.

%Kerr's discovery was quickly generalized to charged rotating black holes
%by Newman et al.\ (1965) \cite{Newman:1965},
%and to other electrovac solutions by Carter (1968) \cite{Carter:1968c}.
%However, no satisfactory interior solutions for rotating black holes
%are known
%interior solution for the Kerr geometry were unsuccessful.
%%\cite{Cohen:1967:Israel:1970,Burinskii:2010kx}.
%\cite{Stephani:2003}.

The papers consider only classical general relativity,
not alternate theories of gravity,
nor speculative quantum processes
that might
%for example destabilize
occur at
the outer horizon.

\penrosekerrinflationfig

The Penrose diagram of the analytically extended Kerr
\cite{Kerr:1963}
geometry,
Figure~\ref{penrosekerrinflation},
provides a good starting point
for understanding where and how the interior Kerr geometry fails.
A spherical charged (Reissner-Nordstr\"om) black hole
has a similar interior structure,
with essentially the same Penrose diagram,
and much of the literature has focussed on this simpler case.
The Kerr geometry, and more generally the Kerr-Newman geometry,
has two inner horizons that are gateways to regions of unpredictability,
signalled by the presence of timelike singularities.
In 1968, Penrose
\cite{Penrose:1968}
pointed out that an observer passing through the outgoing inner horizon
(the Cauchy horizon)
of a spherical charged black hole
would see the outside Universe infinitely blueshifted,
and he suggested that the infinite blueshift would destabilize the inner horizon.
The infinite blueshift is plain from the Penrose diagram,
Figure~\ref{penrosekerrinflation},
which shows that a person passing through the outgoing inner horizon
sees the entire future of the outside Universe go by in a finite time.
Perturbation theory,
much of it expounded in Chandrasekhar's (1983) monograph
\cite{Chandrasekhar:1983},
confirmed that waves from the outside Universe would
amplify to a diverging energy density on the outgoing inner horizon
of a spherical charged black hole.
The result was widely interpreted as indicating the instability
of the inner horizon.

It was not until 1990 that
the full nonlinear nature of the instability at the inner horizon
was eventually clarified by Poisson \& Israel
\cite{Poisson:1990eh}.
Poisson \& Israel
showed that if ingoing %(positive energy)
and outgoing %(negative energy)
streams are simultaneously present just above the inner horizon
of a spherical charged black hole,
then cross-flow between the two streams would lead to an
exponential growth of the interior mass.
They called the instability ``mass inflation.''
Shortly thereafter,
Barrab\`es, Israel \& Poisson \cite{Barrabes:1990}
generalized the arguments to the case of rotating black holes,
showing that whenever two null sheets cross,
an effective mass parameter defined by the product
of the expansions of the null bundles inflates.
The inflationary instability in spherical charged black holes
was confirmed analytically and numerically
in several studies,
as reviewed by \cite{Hamilton:2008zz}.

The physical reason for the inflationary instability
can be seen in the Penrose diagram, Figure~\ref{penrosekerrinflation}.
In the Black Hole region between the outer and inner horizons,
the time coordinate (the one that expresses time-translation symmetry)
is spacelike,
so that it is possible to go either forward or backward in time.
%and consequently to have either positive or negative energy.
Inside the inner horizon, the time coordinate reverts to being timelike.
Ingoing
%positive energy
particles want to fall into a region where
the time coordinate is progressing forwards,
while outgoing
%negative energy
particles want to fall into a region where
the time coordinate is progressing backwards.
The Penrose diagram, Figure~\ref{penrosekerrinflation},
shows that indeed there are two distinct ingoing and outgoing inner horizons
which disgorge on to two causally separated pieces of spacetime
where the time coordinate is pointed in opposite directions.
To achieve this causal separation,
the ingoing and outgoing streams must exceed the speed of light
relative to each other.
This is Penrose's infinite blueshift.
In reality, if ingoing and outgoing streams are present,
then their attempt to exceed the speed of light relative to each other
produces a counter-streaming energy and pressure
that, however tiny the initial streams may be,
inevitably grows to the point that it becomes a significant source of gravity.
As expounded by \cite{Hamilton:2008zz},
the counter-streaming pressure produces a gravitational force
that is in opposite directions for ingoing and outgoing streams,
accelerating the streams ever faster through each other,
in turn increasing the counter-streaming pressure.
The inflationary instability
%is thus self-powered by the gravity of its own counter-streaming,
%and exponential.
thus grows exponentially.

%Besides \cite{Barrabes:1990},
%almost the only prior work on the inflationary instability inside
%rotating black holes is that of Ori
%\cite{Ori:1992,Ori:2001pc}
%(see \cite{Hamilton:2009hu} for more references).
%Ori considered the situation in which
%a black hole undergoes collapse,
%and inflation is driven by Price tails of ingoing and outgoing
%gravitational radiation generated by the collapse.
%The situation is more complicated than that considered
%in the present series of papers,
%and Ori was able to advance only qualitative arguments about the
%character of the instability.

\section{Approach}

The strategy adopted in the present papers
is motivated by two key physical insights.

The first insight is that,
as shown in \S{V} of Paper~2 \cite{Hamilton:2010b},
collisionless ingoing and outgoing streams
falling towards the inner horizon of the Kerr-Newman geometry
become highly focussed into twin narrow, intense beams
pointed along the ingoing and outgoing principal null directions.
The focussing is along these two special directions
regardless of the initial distributions of orbital parameters of the streams.
This implies that the energy-momentum tensor of the ingoing and outgoing streams
takes a simple and predictable form near the inner horizon.
(We use the term inner horizon to describe
the narrow region where inflation takes place,
even though the inner horizon is destroyed by inflation,
and therefore does not actually exist.)

As first shown by
\cite{Carter:1968c},
the Kerr-Newman geometry
(and some other electrovac geometries)
is Hamilton-Jacobi separable in a tetrad aligned with the
principal null directions.
The fact that collisionless streams focus near the inner horizon
along precisely these principal null directions
suggests that the spacetime might continue to be separable
in the presence of inflation.

\kerrrcontourfig

The second insight is that
the geometry of a spherical charged black hole undergoing inflation
at (just above) its inner horizon
has a step-function character.
The spacetime is well-approximated by the electrovac
(Reissner-Nordstr\"om) geometry down to just above the inner horizon.
Then, in a tiny interval of radius and proper time near the inner horizon,
the centre-of-mass counter-streaming energy and pressure,
the Weyl curvature, and the interior mass all inflate
to exponentially huge values.
Counter-intuitively,
the smaller the incident ingoing and outgoing streams,
the more rapidly quantities exponentiate
\cite{Hamilton:2008zz}.
In the limit of tiny accretion rate,
the geometry tends to a step-function.
This suggests that inflationary spacetimes might be found
by looking for solutions with a steplike character.
%The strategy works, and provides a natural explanation
%for why relativists and mathematicians did not find these solutions before.
%The place where the step-function behaviour of inflation
%is entertained is \S{VI\,B} of Paper~2 \cite{Hamilton:2010b}.
The steplike character of the inflationary solutions
can be seen in the sharp turns at the inner horizon
in the contours of constant radius and latitude in Figure~\ref{kerrrcontour}.
%and the contours of constant conformal factor in Figure~\ref{kerrrhocontour}.

As an aside, it is worth commenting on the challenges and pitfalls
of computing inflationary spacetimes numerically
rather than analytically.
One major numerical challenge arises from the fact that
during inflation
physical quantities inflate to exponentially huge values
over tiny intervals of distance and time.
One potential pitfall is that inflation requires
ingoing and outgoing streams that can stream relativistically
through each other.
A code, or indeed analytic model,
that treats the matter as a single fluid with a sound speed
less than the speed of light
%(which includes perfect fluids with $p/\rho < 1$)
artificially suppresses inflation
by disallowing the relativistic counter-streaming that drives it.

\section{Summary derivation}
\label{derivation}

This section summarizes the derivation of the solution
for an uncharged rotating black hole.
Complete details are given in Paper~2.

The solutions presented in this series of papers are
conformally stationary, axisymmetric, and conformally separable.
Conformally stationarity combines the assumption of
conformal time-translation invariance (self-similarity)
with an infinitesimal expansion rate.
Let $x^\mu \equiv \{ x , t , y , \phi \}$ be coordinates
in which $t$ is conformal time,
%with respect to which the spacetime is conformally time-translation symmetric,
$\phi$
is the azimuthal angle,
%with respect to which the spacetime is axisymmetric,
and $x$ and $y$ are radial and angular coordinates.
As shown in Appendix~A of Paper~2 \cite{Hamilton:2010b},
the line-element may be taken to be
%(this repeats equation~(3) of Paper~2 \cite{Hamilton:2010b})
\begin{equation}
\label{lineelement}
  \dd s^2
  =
  \rho^2
  \left[
  {\dd x^2 \over \Deltax}
  -
  {\Deltax \over \sigma^4}
  \left( \dd t - \omegay \, \dd \phi \right)^2
  +
  {\dd y^2 \over \Deltay}
  +
  {\Deltay \over \sigma^4}
  \left( \dd \phi - \omegax \, \dd t \right)^2
  \right]
  \ ,
\end{equation}
where
\begin{equation}
  \sigma
  \equiv
  \sqrt{ 1 - \omegax \, \omegay }
  \ .
\end{equation}
The determinant of the $2 \times 2$ submatrix
of $t$--$\phi$ metric coefficents
defines the radial and angular horizon functions
$\Deltax$ and $\Deltay$:
\begin{equation}
  g_{tt} g_{\phi\phi} - g_{t\phi}^2
  =
  -
  {\rho^4 \over \sigma^4}
  \Deltax \Deltay
  \ .
\end{equation}
Horizons occur when one or other of the horizon functions
$\Deltax$ and $\Deltay$
vanish.
The focus here is the region near the inner horizon
where the radial horizon function $\Deltax$ is negative and tending to zero.
The line-element~(\ref{lineelement}) defines a tetrad
that is aligned with the principal null directions.
Since the radial coordinate $x$ is timelike
near the inner horizon,
it is convenient to take $x$ as the time coordinate of the tetrad,
and to choose the sign of $x$ so that it increases inwards,
the direction of advancing time.

%\kerrrhocontourfig

The conformal factor $\rho$
is a product of separable (electrovac)
$\rhosep$,
time-dependent
$\ee^{\vel t}$,
and
inflationary
$\ee^{-\expinf}$
factors,
\begin{equation}
\label{rho}
  \rho
  =
  \rhosep
  \ee^{\vel t - \expinf}
  \ .
\end{equation}
Conformal time-translation symmetry is expressed by the fact
that the spacetime expands conformally
(that is, without changing shape)
by factor $\rho \rightarrow \ee^{\vel \Delta t} \rho$
when the conformal time increases by $t \rightarrow t + \Delta t$.
Conformal stationarity means taking the limit
of small expansion rate, or small accretion rate,
{\em after\/} calculations are complete,
\begin{equation}
  \vel \rightarrow 0
  \ .
\end{equation}
This is not the same as stationarity,
which sets $\vel$ to zero at the outset.
A feature of inflation is that the smaller the accretion rate,
the faster inflation exponentiates.
Even in the limit of infinitesimal accretion rate,
inflation drives the
centre-of-mass streaming density and pressure,
and the Weyl curvature, to exponentially huge values.
Mathematically,
Einstein's
%and Maxwell's
equations contain terms of order $\sim \vel / \Deltax$
that grow large at the inner horizon $\Deltax \rightarrow -0$
however small the accretion rate $\vel$ may be.

This paper adopts a collisionless fluid as the source of
energy-momentum that ignites and then drives inflation.
Because collisionless particles stream hyper-relativistically
through each other during inflation,
the trajectories of massive freely-falling particles are
well-approximated by those of massless particles.
Conformal separability posits that the equations of motion
of freely-falling massless particles are Hamilton-Jacobi separable,
which implies that a conformal Killing tensor exists.
%For massless particles
%\begin{equation}
%  \pi_t = - E
%  \ , \quad
%  \pi_\phi = L
%  \ .
%\end{equation}
%For massless particles, the Hamilton-Jacobi equation is
%\begin{equation}
%  {\Px^2 - P_t^2
%  \over \Deltax}
%  +
%  {\Py^2 + P_\phi^2
%  \over \Deltay}
%  =
%  0
%  \ ,
%\end{equation}
%where the Hamilton-Jacobi parameters $P_k$ are
%\begin{subequations}
%\label{Pk}
%\begin{align}
%\label{Px}
%  \Px
%  &\equiv
%  - \, \pi_x \Deltax - q \Apot_x
%  %\rho^2
%  %{\dd x \over \dd \lambda}
%  \ ,
%\\
%\label{Pt}
%  P_t
%  &\equiv
%  \pi_t + \pi_\phi \omegax - q \Apot_t
%  \ ,
%\\
%\label{Py}
%  \Py
%  &\equiv
%  \pi_y \Deltay - q \Apot_y
%  %\rho^2
%  %{\dd y \over \dd \lambda}
%  \ ,
%\\
%\label{Pphi}
%  P_\phi
%  &\equiv
%  \pi_\phi + \pi_t \omegay - q \Apot_\phi
%  \ .
%\end{align}
%\end{subequations}
%In terms of $P_k$,
%the covariant tetrad-frame 4-momentum $p_k$
%of the particle is
%\begin{equation}
%\label{pmtetrad}
%  p_k
%  =
%  e_k{}^\kappa
%  \pi_\kappa
%  -
%  q \Apot_k
%  =
%  {1 \over \rho}
%  \left\{
%  {\Px
%  \over \sqrt{- \Deltax}}
%  \ ,
%  {P_t
%  \over \sqrt{- \Deltax}}
%  \ ,
%  {\Py
%  \over \sqrt{\Deltay}}
%  \ ,
%  {P_\phi
%  \over \sqrt{\Deltay}}
%  \right\}
%  \ .
%\end{equation}
%The tetrad-frame covariant electromagnetic 4-potential $A_k$ is
%\begin{equation}
%\label{Apot}
%  A_k
%  \equiv
%  {1 \over \rho}
%  \left\{
%  {\Apot_x \over \sqrt{- \Deltax}}
%  ,
%  {\Apot_t \over \sqrt{- \Deltax}}
%  ,
%  {\Apot_y \over \sqrt{\Deltay}}
%  ,
%  {\Apot_\phi \over \sqrt{\Deltay}}
%  \right\}
%  \ .
%\end{equation}
As shown in Appendix~A of Paper~2,
conformal separability requires that
%The parameters of the line element~(\ref{lineelement})
%and electromagnetic potential~(\ref{Apot})
%the conformal separability conditions
%(these repeat conditions~() of Paper~2 \cite{Hamilton:2010b}).
\begin{equation}
\label{fnxy}
  \begin{array}{ccl}
  \omegax
  \ ,
  &
  \Deltax
  %\ ,
  %&
  %\Apot_x
  %\ ,
  %&
  %\Apot_t
  &
  \mbox{~are functions of $x$ only}
  \ ,
  \\
  \omegay
  \ ,
  &
  \Deltay
  %\ ,
  %&
  %\Apot_y
  %\ ,
  %&
  %\Apot_\phi
  &
  \mbox{~are functions of $y$ only}
  \ .
  \end{array}
\end{equation}
Unlike strict separability (for massive as well massless particles),
conformal separability does not impose
conditions on the conformal factor $\rho$.

Conformally separable inflationary solutions
are obtained by separating the Einstein equations systematically.
Homogeneous solution of the Einstein components
$G_{xy}$, $G_{t\phi}$, and $G_{xx} + G_{tt}$ and $G_{yy} - G_{\phi\phi}$
leads to the usual electrovac solutions
for the electrovac conformal factor $\rhosep$
and the vierbein coefficients $\omegax$ and $\omegay$:
\begin{equation}
\label{rhoxy}
  \rhosep
  =
  \sqrt{
  \rhox^2
  +
  \rhoy^2
  }
  \ , \quad
  \rhox
  =
  \sqrt{
  {g_0 - g_1 \omegax
  \over
  ( f_0 g_1 + f_1 g_0 )
  ( f_0 + f_1 \omegax ) }
  }
  \ , \quad
  \rhoy
  =
  \sqrt{
  {g_1 - g_0 \omegay
  \over
  ( f_0 g_1 + f_1 g_0 )
  ( f_1 + f_0 \omegay )}
  }
  \ ,
\end{equation}
\begin{equation}
\label{domegadvarpi}
  {\dd \omegax \over \dd x}
  =
  2
  \sqrt{
  \left( f_0 + f_1 \omegax \right)
  \left( g_0 - g_1 \omegax \right)
  }
  %( 1 + 2 \rhosep \expinf_r )
  \ , \quad
  {\dd \omegay \over \dd y}
  =
  2
  \sqrt{
  \left( f_1 + f_0 \omegay \right)
  \left( g_1 - g_0 \omegay \right)
  }
  %( 1 + 2 \rhosep \expinf_\theta )
  \ ,
\end{equation}
where $f_0$, $f_1$, $g_0$, and $g_1$ are constants determined
by boundary conditions.
%Inflation modifies the conformal factor $\rho$
%from its electrovac form~(\ref{rhoxy}),
%but leaves intact the equations~(\ref{domegadvarpi})
%governing $\omegax$ and $\omegay$.
Equations~(\ref{rhoxy}) and (\ref{domegadvarpi})
continue to hold throughout inflation and collapse.
The inflationary solutions are generic,
applying wherever a separable electrovac spacetime
has an inner horizon,
so the specific choice of constants
$f_0$, $f_1$, $g_0$, and $g_1$
does not affect the argument.

The most important Einstein equations,
since they lead to equations governing the evolution of
the inflationary exponent $\expinf$
and the horizon function $\Deltax$,
are those for the Einstein components
$G_{xx} - G_{tt}$ and $G_{yy} + G_{\phi\phi}$.
The collisionless source of both these components can be treated
as negligible, in the conformally stationary limit of small accretion rate.
The angular components are negligible because inflation amplifies
the radial, not angular components;
and the trace of the collisionless energy-momentum remains small
because it depends on the rest mass of the particles,
which is unchanged by inflation.
Define $\Ux$, $\Uy$, $\Xx$, $\Xy$, $\Yx$, and $\Yy$ by
\begin{subequations}
\begin{align}
\label{Uxy}
  \Ux
  \equiv
  -
  {\partial \expinf \over \partial x} \Deltax
  \ &, \quad
  \Uy
  \equiv
  {\partial \expinf \over \partial y} \Deltay
  \ ,
\\
\label{Xxy}
  \Xx
  \equiv
  {\partial \Ux \over \partial x}
  +
  2 {\Ux^2 - \vel^2 \over \Deltax}
  \ &, \quad
  \Xy
  \equiv
  {\partial \Uy \over \partial y}
  -
  2 {\Uy^2 + \vel^2 \omegay^2 \over \Deltay}
  \ ,
\\
\label{Yxy}
  \Yx
  \equiv
  {\dd \Deltax \over \dd x}
  +
  3 \Ux
  -
  \Deltax
  {\dd \over \dd x}
  \ln
  \left[ ( f_0 {+} f_1 \omegax ) {\dd \omegax \over \dd x} \right]
  \ &, \quad
  \Yy
  \equiv
  {\dd \Deltay \over \dd y}
  -
  3 \Uy
  -
  \Deltay
  {\dd \over \dd y}
  \ln
  \left[ ( f_1 {+} f_0 \omegay ) {\dd \omegay \over \dd y} \right]
  \ .
\end{align}
\end{subequations}
In terms of these quantities, the Einstein components
$G_{xx} - G_{tt}$ and $G_{yy} + G_{\phi\phi}$ are
\begin{subequations}
\label{GUv}
\begin{align}
\label{GxxmttUv}
  \rho^2 \left( G_{xx} - G_{tt} \right)
  &=
  {1 \over \sigma^2}
  \left(
  \Yx {\dd \ln \omegax \over \dd x}
  -
  \Yy {\dd \ln \omegay \over \dd y}
  \right)
  -
  2 \Xx
  +
  \Yx
  {\dd
  \over \dd x}
  \ln
  \left(
  {f_0 {+} f_1 \omegax \over \omegax}
  \right)
  +
  \Xy
  -
  {\partial \Yy \over \partial y}
  +
  \Yy
  {\dd
  \over \dd y}
  \ln
  \left[
  {\omegay ( f_1 {+} f_0 \omegay ) \over \dd \omegay / \dd y}
  \right]
\nonumber
  \\
  & \quad
  + \,
  \Ux
  {\partial
  \over \partial x}
  \ln
  \left[ \sigma^2 ( f_0 {+} f_1 \omegax ) \right]
  -
  \Uy
  {\partial \over \partial y}
  \ln \left[
  {( g_1 {-} g_0 \omegay ) \over \sigma^2} {\dd \omegay \over \dd y}
  \right]
  \ ,
\\
\label{GyypphiphiUv}
  \rho^2 \left( G_{yy} + G_{\phi\phi} \right)
  &=
  {1 \over \sigma^2}
  \left(
  \Yx {\dd \ln \omegax \over \dd x}
  -
  \Yy {\dd \ln \omegay \over \dd y}
  \right)
  -
  2 \Xy
  -
  \Yy
  {\dd \over \dd y}
  \ln \left( {f_1 {+} f_0 \omegay \over \omegay} \right)
  +
  \Xx
  +
  {\partial \Yx \over \partial x}
  -
  \Yx
  {\dd \over \dd x}
  \ln
  \left[
  {\omegax ( f_0 {+} f_1 \omegax ) \over \dd \omegax / \dd x}
  \right]
\nonumber
  \\
  & \quad
  + \,
  \Uy
  {\partial \over \partial y}
  \ln \left[ \sigma^2 ( f_1 {+} f_0 \omegay ) \right]
  -
  \Ux
  {\partial \over \partial x}
  \ln \left[
  {( g_0 {-} g_1 \omegax ) \over \sigma^2} {\dd \omegax \over \dd x}
  \right]
  \ .
\end{align}
\end{subequations}
Homogeneous solutions of these equations can be found by supposing that
$\Ux$, $\Xx$, and $\Yx$ are all functions of radius $x$,
while
$\Uy$, $\Xy$, and $\Yy$ are all functions of radius $y$,
and by separating each of the equations as
\begin{equation}
\label{sepG}
  {1 \over \sigma^2}
  \left(
  {f_0 h_0 {+} h_2 \omegax {+} f_1 h_1 \omegax^2 \over \omegax}
  -
  {f_1 h_1 {+} h_2 \omegay {+} f_0 h_0 \omegay^2 \over \omegay}
  \right)
  -
  {f_0 h_0 {+} h_3 \omegax \over \omegax}
  +
  {f_1 h_1 {+} h_3 \omegay \over \omegay}
  =
  0
  \ ,
\end{equation}
for some constants $h_0$, $h_1$, $h_2$, and $h_3$.
If one attempts to separate equations~(\ref{GUv}) exactly,
then the attempt fails unless $\Ux$ and $\Uy$ are identically zero,
which is the usual electrovac case.
But if $\Ux$ is taken to be small but finite,
then separation succeeds,
and inflation emerges.
If $\Ux$ and $\Uy$ on the second lines of equations~(\ref{GUv})
are treated as negligibly small,
then separating the first lines of each of equations~(\ref{GUv})
according to the pattern of equation~(\ref{sepG})
leads to the homogeneous solutions
\begin{subequations}
\label{XYxhomog}
\begin{align}
\label{Xxhomog}
  \Xx
  =
  0
  \ &, \quad
  \Xy
  =
  0
  \ ,
\\
\label{Yxhomog}
  \Yx
  =
  {( f_0 + f_1 \omegax ) ( h_0 + h_1 \omegax ) \over \dd \omegax / \dd x}
  \ &, \quad
  \Yy
  =
  {( f_1 + f_0 \omegay ) ( h_1 + h_0 \omegay ) \over \dd \omegay / \dd y}
  \ .
\end{align}
\end{subequations}
If $\Ux = \Uy = 0$,
then solution of equations~(\ref{Yxy}) and (\ref{Yxhomog})
for $\Yx$ and $\Yy$,
subject to appropriate boundary conditions,
yields the radial and angular horizon functions $\Deltax$ and $\Deltay$
of the Kerr line-element.
The result is easily generalized to other electrovac spacetimes
by admitting appropriate sources for $\Yx$ and $\Yy$.

The quantity $\Xy$ defined by equation~(\ref{Xxy})
determines the evolution of $\Uy$,
and the solution $\Xy = 0$, equation~(\ref{Xxhomog}),
then implies that inflation leaves $\Uy$ unchanged,
and sensibly equal to its electrovac value of zero,
in the conformally stationary limit.
Thus inflation leaves the angular horizon function $\Deltay$
unchanged from its electrovac value.

On the other hand, inflation drives $\Ux$ away from zero
however small it might initially be.
In the vicinity of the inner horizon, where $\Deltax \rightarrow -0$,
the solutions~(\ref{XYxhomog}) for $\Xx$ and $\Yx$
defined by equations~(\ref{Xxy}) and (\ref{Yxy})
imply the evolution equations
\begin{subequations}
\label{UDx}
\begin{align}
\label{Ux}
  {\partial \Ux \over \partial x}
  +
  2 {\Ux^2 - \vel^2 \over \Deltax}
  &=
  0
  \ ,
\\
\label{Dx}
  {\dd \Deltax \over \dd x}
  +
  3 \Ux
  &=
  \Deltax^\prime
  \ ,
\end{align}
\end{subequations}
where
$\Deltax^\prime \equiv \left. \dd \Deltax / \dd x \right|_{\xin}$
is the (positive) derivative of the electrovac
horizon function at the inner horizon
$x = \xin$.
Below, equation~(\ref{xinf}),
it will be found that the radius $x$ remains frozen
at its inner horizon value $\xin$ throughout inflation and collapse,
so the right hand side of equation~(\ref{Dx}),
which is the electrovac solution for $\Yx$
evaluated at the inner horizon,
is constant during inflation and collapse.
The evolution equation~(\ref{Ux}) for $\Ux$
involves a term inversely proportional to the horizon function $\Deltax$,
which diverges at the inner horizon $\Deltax \rightarrow -0$,
driving $\Ux$ away from zero
however small $\Ux$ might initially be.

Equation~(\ref{Ux}) leads to instability only at the inner horizon,
where $\Deltax \rightarrow -0$.
At the outer horizon, where $\Deltax \rightarrow 0$,
solutions of equation~(\ref{Ux}) decay rather than grow.

The separation of the Einstein components~(\ref{GUv})
that leads to the evolution equations~(\ref{UDx})
was premised on the assumption that the terms proportional
to $\Ux$ and $\Uy$ on the second lines of equations~(\ref{GUv})
could be neglected.
However,
the separation continues to remain valid during inflation
and collapse when $\Ux$ grows huge.
The reason for this is that the dominant terms
in the Einstein components~(\ref{GUv})
during inflation and collapse
are of order $\Ux^2 / \Deltax$,
coming from the expression~(\ref{Xxy}) for $\Xx$.
Thus,
once $\Ux$ ceases to be negligible,
the condition for the validity of the separation becomes
$\Ux \ll \Ux^2 / | \Deltax |$,
or equivalently
$| \Deltax | \ll \Ux$.
Consequently the condition for the validity of the separation
of the Einstein components~(\ref{GUv}) is
\begin{equation}
\label{sepcondition}
  \mbox{either}
  \quad
  \Ux \ll 1
  \quad
  \mbox{or}
  \quad
  | \Deltax | \ll \Ux
  \ .
\end{equation}
Condition~(\ref{sepcondition})
holds from electrovac through inflation and collapse,
provided that the accretion rate is small,
as conformal stationarity prescribes.
The fact that condition~(\ref{sepcondition}) suffices
is verified in \S{VIII\,J} of Paper~2,
where the Einstein equations are solved to next order in $\Deltax / \Ux$,
and it is shown that the effect on the evolution of the inflationary
exponent $\expinf$ and horizon function $\Deltax$ is negligible.

The evolution equations~(\ref{UDx}) are solved
in the next section, \S\ref{inflationcollapse},
but first it is necessary to attend to the other Einstein equations.

In the conformally separable geometry,
freely-falling collisionless ingoing and outgoing streams
become highly focussed along
the principal ingoing and outgoing null directions
as they approach the inner horizon.
%which lie in the $x$-$t$ plane in the tetrad adopted in this paper.
Inflation accelerates the streams even faster along the same null directions,
causing the $x$ and $t$ components of the tetrad-frame momenta
of freely-falling collisionless streams to grow exponentially.
Consequently the collisionless energy-momentum is dominated by
its $x$-$t$ components.
The associated components of the Einstein tensor are
\begin{equation}
\label{GxtxUv}
  \rho^2
  \left(
  {G_{xx} + G_{tt} \over 2}
  \,\pm\,
  G_{xt}
  \right)
  =
  ( \Ux \mp \vel )
  \left[
  {\Yx \pm \vel \over - \Deltax}
  -
  {\dd \over \dd x}
  \ln \left( \dd \omegax \over \dd x \right)
  \right]
  +
  \Xx
  \ .
\end{equation}
Since $\Xx = 0$,
and $\Yx = \Deltax^\prime$,
and the term proportional to
${\dd \ln ( \dd \omegax / \dd x ) / \dd x}$
is sub-dominant
(in fact the term disappears when the Einstein components~(\ref{GUv})
and corresponding evolution equations~(\ref{UDx})
are solved to next order in $\Deltax / \Ux$; see Paper~2),
equation~(\ref{GxtxUv}) simplifies to
\begin{equation}
\label{GxtxUvsimp}
  \rho^2
  \left(
  {G_{xx} + G_{tt} \over 2}
  \,\pm\,
  G_{xt}
  \right)
  =
  {1 \over - \Deltax}
  ( \Ux \mp \vel )
  ( \Deltax^\prime \pm \vel )
  \ .
\end{equation}
The right hand side of equation~(\ref{GxtxUv}) agrees with $8\pi$
times the energy-momentum tensor of two collisionless streams,
one ingoing ($+$) and one outgoing ($-$),
\begin{equation}
  T_{kl}
  =
  N^+ p^+_k p^+_l
  +
  N^- p^-_k p^-_l
  \ ,
\end{equation}
with densities
\begin{equation}
\label{Npm}
  N^\pm
  =
  {1 \over 16\pi}
  ( \Ux \mp \vel )
  ( \Deltax^\prime \pm \vel )
  \ ,
\end{equation}
and tetrad-frame momenta
\begin{equation}
\label{ppm0}
  p^\pm_k
  =
  {1 \over \rho}
  \left\{
  -
  {1 \over \sqrt{- \Deltax}}
  \, , \ 
  \mp
  {1 \over \sqrt{- \Deltax}}
  \, , \ 
  0
  \, , \ 
  0
  \right\}
  \ .
\end{equation}
The tetrad-frame momenta~(\ref{ppm0}) are null vectors
pointed along the principal ingoing and outgoing null directions.
That equations~(\ref{Npm}) and (\ref{ppm0}) describe correctly
the behaviour of freely-falling streams can be shown by
solving the Hamilton-Jacobi and collisionless Boltzmann equations
(see Paper~2),
and can be confirmed by checking that the densities and momenta
satisfy, to requisite accuracy, covariant number conservation,
$D^k N^\pm p^\pm_k = 0$,
and the geodesic equation
$\dd p^\pm_k / \dd \lambda = 0$,
where $\lambda$ is an affine parameter.
It might seem somewhat miraculous that the
$x$-$t$ components of the Einstein equations
are satisfied with a collisionless source,
but it is no coincidence.
Einstein's equations enforce covariant energy-momentum conservation,
$D^k T_{kl} = 0$.
Since the angular components are sub-dominant,
only the 3 distinct $x$-$t$ components of the energy-momentum tensor
are important.
The 3 components are subject to 2 energy-momentum conservation equations,
but in the present instance the 2 conservation equations are redundant,
so there is effectively 1 conservation equation.
However, the energy-momentum conservation equations for freely-falling
ingoing and outgoing streams are symmetrically related to each other by
$\vel \rightarrow - \vel$,
so conservation of the sum of their energies,
as enforced by Einstein,
implies conservation of both.
The two conservation equations,
coupled with solution of the Einstein equation for
$G_{xx} - G_{tt}$, equation~(\ref{GxxmttUv}),
leads to a complete, self-consistent set of equations.

The angular motions of the freely-falling streams are small
compared to their radial motions, but not necessarily zero.
Next in order of magnitude,
after the 3 radial ($x$-$t$) components
of the energy-momentum tensor,
are its 4 off-diagonal radial-angular components.
The corresponding components of the Einstein tensor are
\begin{subequations}
\label{GxtyphiUv}
\begin{align}
\label{GxtyUv}
  \rho^2
  \left(
  G_{xy}
  \,\pm\,
  G_{ty}
  \right)
  &=
  -
  {1 \over \sqrt{- \Deltax \Deltay}}
  ( \Ux \mp \vel )
  \left(
  \Deltay
  {\partial \ln \rhosep^2 \over \partial y}
  -
  2 \Uy
  \right)
  -
  \sqrt{- \Deltax \over \Deltay}
  \left(
  \Uy
  {\partial \ln \rhosep^2 \over \partial x}
  \, \pm \,
  {\vel \omegay \over \sigma^2} {\dd \omegay \over \dd y}
  \right)
  %+
  %{1 \over 2}
  %\sqrt{ {\Deltay \over - \Deltax} }
  %{\partial \Yx \over \partial y}
  \ ,
\\
\label{GxtphiUv}
  \rho^2
  \left(
  G_{x\phi}
  \,\pm\,
  G_{t\phi}
  \right)
  &=
  \pm
  {1 \over \sqrt{- \Deltax \Deltay}}
  ( \Ux \mp \vel )
  \left(
  {\Deltay \over \sigma^2}
  {\dd \omegax \over \dd x}
  \mp
  2 \vel \omegay
  \right)
  \,\mp\,
  \sqrt{- \Deltax \over \Deltay}
  \left(
  {\Uy \over \sigma^2} {\dd \omegay \over \dd y}
  \mp
  \vel \omegay {\partial \ln \rhosep^2 \over \partial x}
  \right)
  \ .
\end{align}
\end{subequations}
Since $\Uy$
%and $\partial \Yx / \partial y = 0$,
and the terms proportional to $\vel \sqrt{- \Deltax}$ are negligible,
equations~(\ref{GxtyphiUv}) simplify to
\begin{subequations}
\label{GxtyphiUvsimp}
\begin{align}
\label{GxtyUvsimp}
  \rho^2
  \left(
  G_{xy}
  \,\pm\,
  G_{ty}
  \right)
  &=
  -
  {1 \over \sqrt{- \Deltax \Deltay}}
  ( \Ux \mp \vel )
  \Deltay
  {\partial \ln \rhosep^2 \over \partial y}
  \ ,
\\
\label{GxtphiUvsimp}
  \rho^2
  \left(
  G_{x\phi}
  \,\pm\,
  G_{t\phi}
  \right)
  &=
  \pm
  {1 \over \sqrt{- \Deltax \Deltay}}
  ( \Ux \mp \vel )
  \left(
  {\Deltay \omegax^\prime \over \sigma^2}
  \mp
  2 \vel \omegay
  \right)
  \ ,
\end{align}
\end{subequations}
where
$\omegax^\prime \equiv \left. \dd \omegax / \dd x \right|_{\xin}$,
which is effectively constant throughout inflation and collapse,
is the derivative of $\omegax$ at the inner horizon $x = \xin$,
equation~(\ref{domegadvarpi}).
The right hand sides of equations~(\ref{GxtyphiUvsimp})
agree with $8\pi$ times the energy-momentum tensor of
ingoing and outgoing streams with the same densities $N^\pm$
as before, equation~(\ref{Npm}), but with tetrad-frame momenta
$p^\pm_k$
having finite rather than zero angular components:
\begin{equation}
\label{ppm}
  p^\pm_k
  =
  {1 \over \rho}
  \left\{
  -
  {1 \over \sqrt{- \Deltax}}
  \, , \
  \mp
  {1 \over \sqrt{- \Deltax}}
  \, , \
  {1 \over \sqrt{\Deltay}}
  \left(
  {\Deltay
  \partial \ln \rhosep^2 / \partial y
  \over
  \Deltax^\prime \pm \vel}
  \right)
  \, , \
  \mp
  {1 \over \sqrt{\Deltay}}
  \left(
  {\Deltay \omegax^\prime / \sigma^2 \mp 2 \vel \omegay
  \over
  \Deltax^\prime \pm \vel}
  \right)
  \right\}
  \ .
\end{equation}
The tetrad-frame momenta~(\ref{ppm})
satisfy the Hamilton-Jacobi equations
with constant Hamilton-Jacobi parameters
along the path of the streams.
The angular components of the momenta
are small compared to the radial components as long as
\begin{equation}
\label{Dsmall}
  | \Deltax | \ll 1
  \ .
\end{equation}
The momentum~(\ref{ppm}) is hyper-relativistic,
and $p^\pm_t = \pm p^\pm_x$ to an excellent approximation
so long as condition~(\ref{Dsmall}) is true.
If the condition~(\ref{Dsmall}) is violated,
then it signifies that angular motions are becoming important,
and the solution is breaking down.

It should be emphasized that,
as long as condition~(\ref{Dsmall}) holds,
the purely radial ($x$-$t$) Einstein equations hold regardless of
angular motions,
and thus the radial solution is unaffected by angular motions.
However,
if one requires that the sub-dominant radial-angular Einstein equations
are also satisfied,
then the angular motion of the collisionless streams must be
as given by equation~(\ref{ppm}).
One might perhaps have expected that conformally separable solutions
would require that the collisionless streams
would move exactly along the principal null directions,
but equation~(\ref{ppm}) shows that this is not true.
%The fact that some motion is required in the angular $y$ direction (latitude)
%can be attributed to the fact that surfaces of constant conformal factor $\rho$
%do not coincide with surfaces of constant radius $x$.

Again, it might seem remarkable that the radial-angular Einstein equations
are satisfied by collisionless ingoing and outgoing streams.
And again, this coincidence results from energy-momentum conservation.
There are 4 radial-angular Einstein components,
subject to 2 energy-momentum conservation equations.
The energy-momentum conservation equations for the freely-falling
ingoing and outgoing
streams are symmetrically related by $\vel \rightarrow - \vel$,
so conservation of their sum implies conservation of both.

The final, sub-sub-dominant, components of the energy-momentum tensor
are the 3 purely angular ($y$-$\phi$) components.
The component $G_{yy} + G_{\phi\phi}$ component has already been addressed,
equation~(\ref{GyypphiphiUv}).
The remaining 2 components are
\begin{equation}
\label{GyphiyUv}
  \rho^2
  \left(
  {G_{yy} - G_{\phi\phi} \over 2}
  \,\pm\,
  \im
  G_{y\phi}
  \right)
  =
  ( \Uy \mp \im \vel \omegay )
  \left[
  {- \,
  \Yy
  \pm
  \im \vel \omegay \over \Deltay}
  -
  {\dd \over \dd y}
  \ln \left( \dd \omegay \over \dd y \right)
  \right]
  +
  \Xy
  \,\mp\,
  \im \vel {\dd \omegay \over \dd y}
  \ .
\end{equation}
Since $X_y = 0$,
and $U_y$ and $\vel$ are negligibly small,
and there are no denominators of the radial horizon function $\Deltax$,
equation~(\ref{GyphiyUv}) simplifies to
\begin{equation}
\label{GyphiyUvsimp}
  \rho^2
  \left(
  {G_{yy} - G_{\phi\phi} \over 2}
  \,\pm\,
  \im
  G_{y\phi}
  \right)
  =
  0
  \ .
\end{equation}
During inflation,
the collisionless streams have negligible angular components
of energy-momentum
because the densities of the accreting streams are negligible,
in the conformally stationary limit,
and inflation does not amplify angular motions.
During collapse, the conformal factor $\rho$ shrinks,
and angular motions grow.
However, as long as the angular motions are sub-dominant,
which is true as long as condition~(\ref{Dsmall}) is satisfied,
the angular components of the energy-momentum can be neglected consistently:
the Einstein equations for the purely radial components,
and for the radial-angular components,
are unaffected by the angular components~(\ref{GyphiyUvsimp})
(the angular component $G_{yy} + G_{\phi\phi}$,
along with $G_{xx} - G_{tt}$,
equations~(\ref{GUv}),
determined the angular horizon function $\Deltay$,
which inflation leaves unaltered from its electrovac form).
Equation~(\ref{GyphiyUvsimp}) requires that the $2 \times 2$
angular submatrix of the energy-momentum tensor be isotropic,
proportional to the $2 \times 2$ unit matrix.
As discussed in Paper~2,
it is possible to arrange the angular energy-momentum to be isotropic
by admitting multiple ingoing and outgoing streams,
with mean momenta set by equation~(\ref{ppm})
and isotropic mean squared momenta.
Treating the diagonal components of the angular energy-momentum
requires taking equations~(\ref{GUv}) to next order in $\Deltax / \Ux$,
but this can be done.

Eventually however,
the angular components do become important,
when $| \Deltax | \sim 1$,
and the solution fails.

\inflationaryhorizonfig

\section{Inflation and collapse}
\label{inflationcollapse}

Denote the initial value of $\Ux$, equation~(\ref{Uxy}),
incident on the inner horizon by $\uel$,
a small parameter of order $\vel$,
\begin{equation}
  \Ux = \uel
  \quad
  \mbox{initially}
  \ .
\end{equation}
The densities $N^\pm$ of ingoing and outgoing streams
incident on the inner horizon are proportional to $\uel \mp \vel$,
equation~(\ref{Npm}).
Inflation is driven by counter-streaming
between ingoing and outgoing streams,
so both streams must be present for inflation to occur,
but even the tiniest amount suffices to trigger inflation.
Positivity of both ingoing and outgoing densities requires that
\begin{equation}
  \uel > \vel > 0
  \ ,
\end{equation}
the condition $\vel > 0$ coming from the fact that the black hole
must expand as it accretes.
The case $\vel = 0$ is the stationary (or homogeneous)
approximation  of
\cite{Burko:1997xa}.
The densities $N^\pm$, equation~(\ref{Npm}),
are also proportional to $\Deltax^\prime \mp \vel$,
where $\Deltax^\prime$ is the positive derivative of the electrovac
horizon function at the inner horizon.
Positivity of both ingoing and outgoing densities
requires $\Deltax^\prime$ to be strictly positive,
which excludes extremal black holes,
whose inner and outer horizons coincide,
and which have $\Deltax^\prime = 0$ at the horizon.

Solution of the evolution equations~(\ref{UDx})
for $\Ux$ and $\Deltax$ yields
\begin{subequations}
\label{UDxinf}
\begin{align}
  \Ux
  &=
  \sqrt{
  \vel^2
  +
  ( \uel^2 - \vel^2 ) 
  \ee^{4 \expinf}
  }
  \ ,
\\
  \Deltax
  &=
  -
  \left( {\Ux^2 - \vel^2 \over \uel^2 - \vel^2} \right)^{3/4}
  \left[
  {
  ( \Ux + \vel )
  ( \uel - \vel )
  \over
  ( \Ux - \vel )
  ( \uel + \vel )
  }
  \right]^{\Deltax^\prime / ( 4 \vel )}
  \ ,
\\
\label{xinf}
  x - \xin
  &=
  -
  \int
  {\Deltax \, \dd \Ux \over 2 ( \Ux^2 - \vel^2 )}
  \ .
\end{align}
\end{subequations}
The integral on the right hand side of equation~(\ref{xinf})
can be expressed analytically as an incomplete beta function,
but the expression is not useful.
Physically, equation~(\ref{xinf}) says
that the radius $x$ is frozen at its inner horizon value $\xin$
during inflation and collapse,
where $\Ux$ is growing, while $\Deltax$ remains small.

Figure~\ref{inflationaryhorizon}
illustrates the evolution of $\Ux$ and the horizon function $\Deltax$
as a function of the inflationary exponent $\xi$,
for parameters $\vel = 0.001$ and $\uel = 0.002$.
The value $\vel = 0.001$,
which physically represents the velocity
with which a distant observer sees the characteristic radius
of the black hole expand,
is large compared to a typical astronomical accretion rate,
but a large value is needed to avoid numerical overflow.
Inflation ignites near the inner horizon
as the horizon function $| \Deltax |$ drops below $\uel$.
During inflation,
the horizon function $| \Deltax |$ decreases exponentially,
while $\Ux$ increases slowly.
During inflation, the inflationary exponent $\expinf$
in the conformal factor $\rho$, equation~(\ref{rho}),
satisfies
\begin{equation}
\label{expinfderivs}
  \expinf
  \ll {\dd \expinf \over \dd x}
  \ll {\ddsq \expinf \over \dd x^2}
  \ ,
\end{equation}
which is the behaviour characteristic of a step-function.
The inequalities~(\ref{expinfderivs}) essentially say
that the acceleration $\ddsq \expinf / \dd x^2$
of the inflationary exponent, which is driven by the
radial energy-momentum of the collisionless streams,
is much larger than the velocity $\dd \expinf / \dd x$,
which in turn is much larger than the distance moved $\expinf$.

Inflation ends when
the absolute value of the horizon function reaches a minimum,
at an exponentially tiny value,
\begin{equation}
  | \Deltax | \sim \ee^{- 1 / \vel}
  \ ,
\end{equation}
at which point the spacetime collapses.
During collapse,
the horizon function $| \Deltax |$ increases,
while the conformal factor
$\rho \propto \ee^{- \expinf}$
shrinks exponentially,
no longer satisfying the inequalities~(\ref{expinfderivs}).
The radial coordinate $x$ remains frozen
even while the conformal factor $\rho$ is shrinking.
That the spacetime collapses rather than leads to a null singularity
accords with the conclusion of
\cite{Hamilton:2008zz}
that the outcome of inflation is collapse
when a black hole continues to accrete,
as is ensured in the present case by the assumption of conformal time
translation invariance (self-similarity).

During collapse the horizon function increases back to of order unity,
$| \Deltax | \sim 1$.
At this point the angular motion of the freely-falling ingoing and outgoing
streams becomes comparable to their radial motion,
and the solution breaks down.
This happens when the conformal factor has collapsed to an exponentially
tiny value,
\begin{equation}
  \rho \sim \ee^{- 1 / \vel}
  \ .
\end{equation}

The right panel of Figure~\ref{inflationaryhorizon}
shows the magnitudes of the radial, radial-angular, and angular components
$T_{xx}$, $T_{xy}$, and $T_{yy}$
of the tetrad-frame energy-momenta of the collisionless streams.
During inflation,
the radial energy-momentum grows fastest,
reaching an exponentially huge value
\begin{equation}
  T_{xx} \sim \ee^{1 / \vel}
  \ .
\end{equation}
During collapse,
the angular energy-momentum grows fastest.

The Weyl curvature tensor has only a spin-$0$ component,
which classifies the spacetime as Petrov Type~D.
The right panel of Figure~\ref{inflationaryhorizon}
shows minus the polar (real) part
of the spin-$0$ component of the Weyl curvature.

It is notable that
the smaller the accretion rate $\vel$,
the more rapidly inflation exponentiates,
and the larger the energy-momentum and curvature grow,
in agreement with the conclusions of
\cite{Hamilton:2004aw,Hamilton:2008zz}.

\section{Boundary conditions}

The solutions are determined by boundary conditions of the
collisionless streams incident on the inner horizon.
Since the solution above the inner horizon is well-approximated
by the Kerr (or other electrovac) solution,
the behaviour of gas above the inner horizon does not affect the solution.

The requirement of conformal separability imposes special boundary conditions.
The densities $N^\pm$ of ingoing ($+$) and outgoing ($-$)
streams incident on the inner horizon are, equation~(\ref{Npm}),
since $\Ux = \uel$ initially,
\begin{equation}
\label{Npminit}
  N^\pm
  =
  {1 \over 16\pi}
  ( \uel \mp \vel )
  ( \Deltax^\prime \pm \vel )
  \ .
\end{equation}
This is just a constant, independent of angular position on the inner horizon.
Thus conformal separability requires that the incident flow
of ingoing and outgoing streams be ``monopole,''
independent of latitude.
It makes physical sense that conformal separability
would require this high degree of symmetry of the accretion flow.

As emphasized in \S\ref{derivation},
because the radial motions of collisionless streams dominate
their angular motions during inflation and collapse
(up until the angular motions become important, at $| \Deltax | \sim 1$),
the radial Einstein equations are unaffected
by the angular motion, and the boundary condition~(\ref{Npminit})
is all that is needed to ensure conformal separability
with sufficient accuracy.
However, if it is required that the sub-dominant radial-angular
components of the Einstein equations are also satisfied,
which is a more stringent constraint on conformal separability,
then the tetrad-frame momenta $p^\pm_k$ of the streams
must have finite angular components,
satisfying equation~(\ref{ppm}).
Figure~\ref{rotinforbit} illustrates the required flow pattern
for a black hole of spin parameter $a = 0.96 \Mbh$.
The energy per unit mass of infalling particles,
which is unspecified by boundary conditions,
is chosen here to be $E/m = \pm 1$.

\rotinforbitfig

As described in Paper~2,
the required angular flow pattern cannot be achieved with collisionless
streams falling from outside the outer horizon.
Streams that fall from outside the outer horizon
must necessarily be ingoing at the outer horizon,
which eliminates half the phase space available to collisionless streams,
making it impossible to satisfy the required conditions
on the angular motion of the streams.
Thus, if the angular conditions are imposed,
then the collisionless ingoing and outgoing streams must be considered
as being delivered ad hoc to just above the inner horizon.
%This is a defect, but not fatal.
%Firstly, the collisionless assumption is likely to fail above the
%inner horizon, where baryons are collisional.
%Secondly,
%the angular conditions are not essential to 
%the radial part of the solution,
%which  holds provided that the monopole condition~(\ref{Npminit}) is satisfied.
%If the angular conditions are relaxed,
%then the required boundary conditions~(\ref{Npminit})
%can easily be satisfied by collisionless streams
%falling from outside the outer horizon.
%Although particles are necessarily ingoing at the outer horizon,
%they can switch from ingoing to outgoing between the inner and outer horizons
%if they have a large enough angular momentum
%in the same direction as the black hole.

\section{Conclusion}

This paper has presented conformally stationary, axisymmetric,
conformally separable solutions for the interior structure of an uncharged
rotating black hole
that undergoes inflation just above its inner horizon, then collapses.
It has long been known that linear perturbations diverge at
the inner horizon of the Kerr geometry,
and it has been suspected that perturbations
would develop nonlinearly similarly to the inflationary instability
\cite{Poisson:1990eh}
known to operate in spherical charged black holes
\cite{Hamilton:2008zz},
and expected to occur also in rotating black holes
\cite{Barrabes:1990}.
The self-consistent nonlinear solutions found here
confirm that,
at least in the conformally separable special case considered here,
the inflationary instability develops
in rotating black holes as anticipated.
%The solutions in this paper are special,
%requiring that the accretion flow incident on the inner horizon
%take a certain precise form in order to maintain separability.
%Nevertheless,
%the generic and inevitable character
%of the inflationary instability
%suggests that the solutions here may be prototypical
%of what happens in real astronomical black holes.

A feature of the Kerr geometry (and other separable electrovac geometries)
is that, as
freely-falling ingoing and outgoing particles approach the inner horizon,
they become highly focussed along the ingoing and outgoing null directions,
regardless of their initial angular motion.
However small the accretion rate may be,
eventually the energy and pressure of the twin beams of particles
counter-streaming hyper-relativistically along the ingoing and outgoing
null directions grows large enough to be a source of gravity.
The gravity produced by the counter-streaming acts to accelerate
ingoing and outgoing stream even faster through each other,
leading to an exponential growth in the streaming density and pressure,
and in the curvature.
This is inflation.
The huge gravitational acceleration produced by the counter-streaming
is in the inward direction, to smaller radius,
but each stream thinks that they are moving in the inward direction,
so the streams are accelerated in opposite directions.

Inflation takes place over an extremely short interval of proper time.
Inflation is like a bullet fired in the chamber of a gun:
an explosion accelerates the bullet,
and shortly after the bullet achieves high velocity,
but still the bullet has hardly moved
(see the inequalities~(\ref{expinfderivs})).
Inflation does in due course alter the geometry,
but in a predictable way:
the conformal factor,
having been accelerated to huge velocity inward,
proceeds to shrink rapidly.
The geometry collapses.

During inflation,
the ingoing and outgoing streams were accelerated along the principal
null directions, without amplifying the angular motion.
During collapse, the angular motions grow.
At an exponentially tiny scale,
the angular motions become comparable to the radial motion,
and the solutions considered in this paper break down.
What happens then is undetermined.

The existence of conformally separable solutions for
the inflationary zones of rotating black holes is not surprising.
Ingoing and outgoing streams focus along
the principal null directions as they approach the inner horizon,
and the streaming energy and pressure generated by the radial beams
accelerates the streams along the same null directions.
The acceleration depends on the accretion rate.
%being faster for smaller accretion rates.
If the densities of ingoing and outgoing streams incident on
the inner horizon are uniform, independent of latitude,
then inflation accelerates the beams at the same rate at all latitudes.
When the geometry begins to collapse, it does so uniformly,
preserving conformal separability.
What happens in the more general case when the accretion flow
on to the inner horizon varies with angular position
remains to be seen.
But that inflation will occur is physically inevitable.

\begin{acknowledgements}
This work was supported by NSF award
AST-0708607.
\end{acknowledgements}

\section*{References}

\bibliographystyle{unsrt}
\bibliography{bh}

\end{document}